\documentclass[sigconf, balance]{acmart}

\def\BibTeX{{\rm B\kern-.05em{\sc i\kern-.025em b}\kern-.08emT\kern-.1667em\lower.7ex\hbox{E}\kern-.125emX}}
    
\copyrightyear{2019}
\acmYear{2019}
\setcopyright{acmcopyright}
\acmConference[MM '19]{Proceedings of the 27th ACM International Conference on Multimedia}{October 21--25, 2019}{Nice, France}
\acmBooktitle{Proceedings of the 27th ACM International Conference on Multimedia (MM '19), October 21--25, 2019, Nice, France}
\acmPrice{15.00}
\acmDOI{10.1145/3343031.3350869}
\acmISBN{978-1-4503-6889-6/19/10}
\setcitestyle{numbers,sort&compress}
\begin{document}
\fancyhead{}
%
\title[Focus Your Attention]{Focus Your Attention: A Bidirectional Focal Attention Network for Image-Text Matching}

%
\author{Chunxiao Liu}
\orcid{0000-0002-3307-9558}
\affiliation{%
  \institution{Institute of Information Engineering, Chinese Academy of Sciences\\
  	School of Cyber Security, University of Chinese Academy of Sciences}
}
\email{liuchunxiao@iie.ac.cn}

\author{Zhendong Mao}
\authornote{Zhendong Mao is the corresponding author.}
\affiliation{%
  \institution{University of Science and \\ Technology of China}}
\email{maozhendong2008@gmail.com}

\author{An-An Liu}
\affiliation{%
  \institution{Tianjin University}}
\email{ana0422@gmail.com}

\author{Tianzhu Zhang}
\affiliation{%
 \institution{University of Science and \\ Technology of China}}
\email{tzzhang10@gmail.com}
 
\author{Bin Wang}
\affiliation{%
  \institution{Xiaomi AI Lab}}
\email{wangbin11@xiaomi.com}

\author{Yongdong Zhang}
\affiliation{%
  \institution{University of Science and \\ Technology of China}}
\email{zhyd73@ustc.edu.cn}

%
\renewcommand{\shortauthors}{Liu and Mao, et al.}
%
\begin{abstract}
Learning semantic correspondence between image and text is significant as it bridges the semantic gap between vision and language. The key challenge is to accurately find and correlate shared semantics in image and text. Most existing methods achieve this goal by representing the shared semantic as a weighted combination of all the fragments (image regions or text words), where fragments relevant to the shared semantic obtain more attention, otherwise less. However, despite relevant ones contribute more to the shared semantic, irrelevant ones will more or less disturb it, and thus will lead to semantic misalignment in the correlation phase. To address this issue, we present a novel Bidirectional Focal Attention Network (BFAN), which not only allows to attend to relevant fragments but also diverts all the attention into these relevant fragments to concentrate on them. The main difference with existing works is they mostly focus on learning attention weight while our BFAN focus on eliminating irrelevant fragments from the shared semantic. The focal attention is achieved by preassigning attention based on inter-modality relation, identifying relevant fragments based on intra-modality relation and reassigning attention. Furthermore, the focal attention is jointly applied in both image-to-text and text-to-image directions, which enables to avoid preference to long text or complex image. Experiments show our simple but effective framework significantly outperforms state-of-the-art, with relative Recall@1 gains of 2.2\% on both Flicr30K and MSCOCO benchmarks.
\end{abstract}

%
%


%

%

%
\maketitle

\section{Introduction}
There is a surge of interest in image-text matching since it bridges the semantic gap between vision and language, which has the potential to integrate multimodal information into existing applications such as the search engine, recommendation system and question answering system. The key challenge in image-text matching is to accurately find and associate shared semantics in image and text. 

\begin{figure}[!t]
	\centering
	\setlength{\abovecaptionskip}{1pt}%
	\setlength{\belowcaptionskip}{0pt}%
	\includegraphics[width=\linewidth]{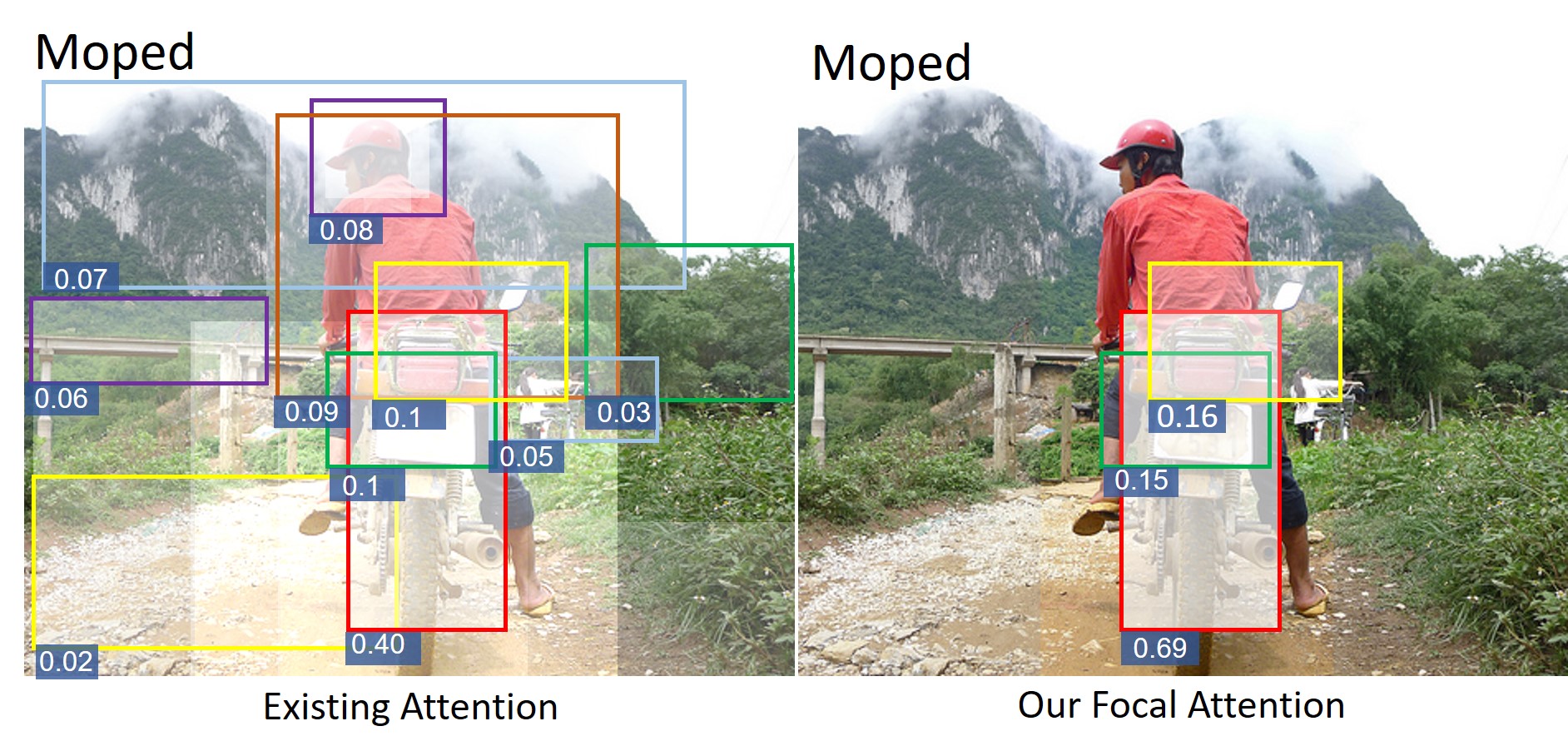}
	\caption{Existing attention vs focal attention. The attended regions are bounded by color box, where whiteness reflects attention weight. Existing attention attends to regions irrelevant to text query ``moped'', like road, bridge and mountain, which will lead to semantic misalignment as ``moped'' is learned to be similar to irrelevant regions. The focal attention avoids it by eliminating irrelevant regions.}
	\Description{intro.}
\end{figure}

Existing image-text matching approaches focus on learning a neural network to find and associate shared semantics in image-text pairs. Early works \cite{ Liu2017LearningAR,Ma2015Multimodal,Niu2017HierarchicalML} achieve this goal by projecting all the fragments (regions and words) in image and text into a latent space without using attention mechanism. Motivated by recent progress in other cross-modal applications like visual question answering and image caption \cite{baltruvsaitis2019multimodal,xu2015show,bahdanau2014neural,yu2017multi,seo2016bidirectional,yang2016stacked,Lu2016Hierarchical}, attention has become an important component in image-text matching framework as it allows to look less on unimportant fragments and look more on important fragments in terms of specific semantic. Some attention-based approaches attend to fragments from different modalities parallelly. A typical approach is \cite{nam2017dual} that separately performs multi-step attention operation in image and text branch, in which all the shared semantics can be discovered step-by-step. Several extensions have been presented, \cite{huang2017instance} holds the insight that partial regions and words contribute to the global semantic. They propose to iteratively select important region-word pairs and learn to maximumly associate them. Despite much progress has been achieved by the above models, they neglect that the importance of regions is dynamically changed with respect to different words, and the importance of words is also dynamically changed with respect to different regions. To solve this problem, \cite{lee2018stacked} proposes to attend to fragments interactively. They develop a more interpretable framework that determines the attention of fragments based on fragments from another modality. Similar works \cite{ Xu_2018_CVPR, huang2019bi} have been proposed motivated by the above model. Nonetheless, these approaches follow an invariant attention framework in which shared semantics are discovered by attending differentially over all the fragments.

However, despite that shared semantics can be found in previous attention mechanism, they cannot be reflected accurately. It is because many fragments are irrelevant to shared semantics, which are also attended, and thus shared semantics will be more or less disturbed. As a result, it will lead to semantic misalignment when learning to associate the shared semantics selected from image and text, i.e., irrelevant fragments from different modalities being closely correlated except for relevant fragments. As is illustrated in Figure 1, given a text fragment ``moped'', conventional attention methods not only attend to the target image region but also attend to its irrelevant regions like road and mountain, which will incorrectly improve relevance between ``moped'' and these irrelevant regions while training. Consequently, only attending to relevant fragments is crucial for learning accurate region-word correspondence.

In this paper, we propose a novel Bidirectional Focal Attention Network (BFAN) to address the semantic misalignment by only attending to relevant fragments instead of all the fragments. This is in contrast to traditional attention where the focal attention focus on irrelevant fragments removal, such that the shared semantics selected from image and text are highly relevant. The focal attention is achieved by preassigning attention, identifying relevant fragments and reassigning attention. Though it is hard to identify relevant fragments without explicit annotation, the focal attention is able to find them by learning a function that scores each fragment based on its preassigned attention relative to other fragments. Fragments that obtain higher preassigned attention than most other fragments with high confidence will be considered as relevant, otherwise irrelevant. The intuition behind this strategy is the attention distribution can roughly determine the gap in relevant and irrelevant fragments. Furthermore, we maximumly associate image-text pairs by applying the focal attention into both image-to-text and text-to-image directions as it avoids preference to long text or complex image. 

The major contributions of this work can be summarized as 
(1) We propose a novel Bidirectional Focal Attention Network that can learn semantic alignment accurately by only focusing on relevant fragments. The focal attention is presented to score each fragment based on relative attention to all other fragments. To the best of our knowledge, it is the first work that only attends to relevant fragments and ignores irrelevant fragments in image-text matching.
(2) We jointly integrate image-to-text and text-to-image matching into a unified framework, which enables to avoid the preference to long text or complex image and maximumly associate relevant image-text pairs.
(3) We conduct extensive experiments on benchmarks, which demonstrate the proposed simple bidirectional focal attention network significantly outperforms state-of-the-art.

\begin{figure*}[!t]
	\centering
	\setlength{\abovecaptionskip}{2pt}%
	\setlength{\belowcaptionskip}{0pt}%
	\includegraphics[width=0.9\linewidth]{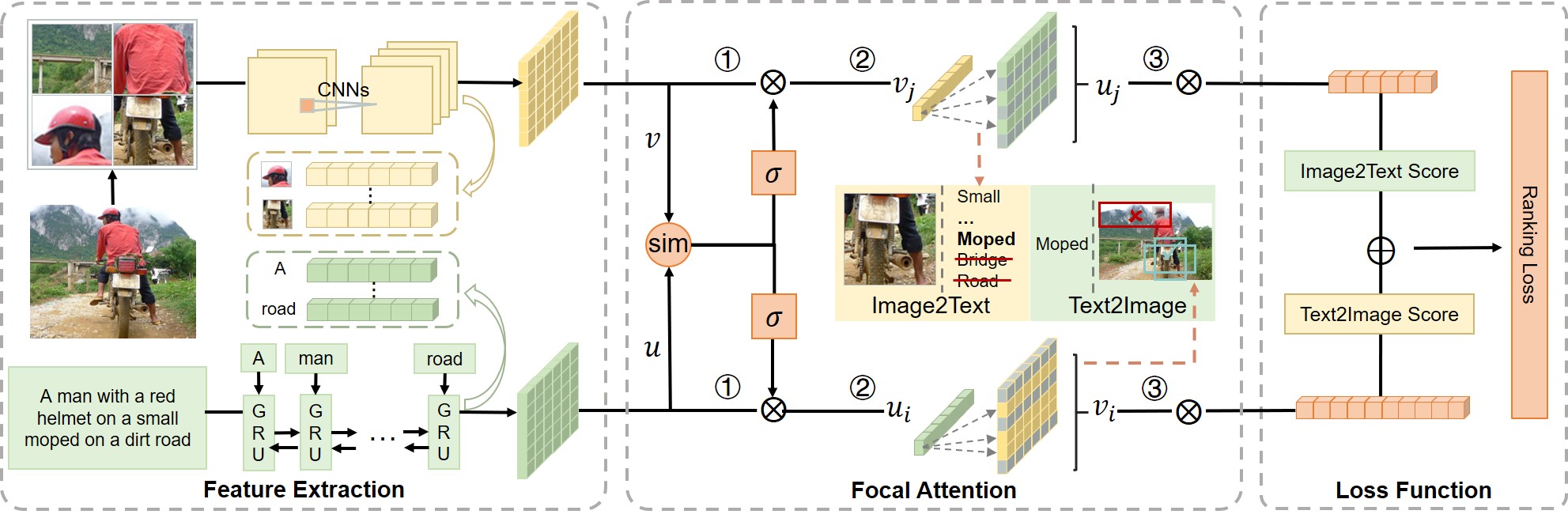}
	\caption{The overall framework of BFAN that consists of feature extraction, focal attention and loss function module. The focal attention module takes the extracted feature as input, and then attends to regions and words interactively. Specifically, \textcircled{\small{1}}it preassigns attention to all the fragments in one modality based on fragments from another modality; \textcircled{\small{2}}it identifies partial relevant parts based on internal relationship of fragments within the same modality; \textcircled{\small{3}}it reassigns the focal attention at image2text and text2image directions, which is then jointly integrated to be optimized by pair-wise ranking loss.}
	\Description{The overall framework of bidirectional focal attention network (BFAN).}
\end{figure*}

\section{Related Work}
Many approaches have been proposed for image-text matching, which can be roughly grouped into one-to-one and many-to-many approaches. The one-to-one approaches learn correspondence between the whole image and text, and the many-to-many approaches learn correspondence between image regions and text words. 

A general solution for one-to-one matching is to associate shared semantics in image-text pairs by projecting them into a common space and optimizing their relevance. Many works \cite{zhang2018deep,zheng2017dual} associate shared semantics by improving multimodal representations. One of the most typical works is \cite{Kiros} that makes the first attempt to encode image and text using Convolution Neural Networks and Recurrent Neural Networks. Similar works are proposed to associate semantic representations in common space by integrating different modules, such as identity mapping \cite{Liu2017LearningAR}, character-level inception \cite{ wehrmann2018bidirectional} and generative adversarial networks \cite{ Gu2017Look}. Different from them, \cite{Eisenschtat2017Linking, karpathy2015deep, huang2018learning, wu2018learning,wang2019learning, wu2017online} learn to associate shared semantics using different objective functions. One of the most popular function is triplet ranking loss \cite{karpathy2015deep} that forces relevant image-text pairs being more similar than irrelevant pairs by a fixed margin. \cite{huang2018learning} goes a step further by attending to hard negatives in the ranking loss function, which significantly improves the performance. Some \cite{wu2018learning,wang2019learning} restrain the multimodal representations by preserving neighborhood structure or geometric structure. Inspired by recent progress achieved by attention mechanism, several attention-based image-text matching methods have been proposed, because it enables to attend to features differentially based on their contribution to shared semantics. A typical work is \cite{nam2017dual} that finds shared semantics by attending to specific vectors from image and text. Many extensions have also been proposed, like \cite{Fan2018StackedLA,Li2017IdentityAwareTM}.

Many-to-many approaches usually learn a latent region-word correspondence through correlating shared semantics comprised of regions and words. It is first proposed by Karpathy et al. \cite{ DBLP:journals/corr/KarpathyJF14} that selects shared semantics by finding most similar region-word pairs. Following this idea, \cite{ Niu2017HierarchicalML} presents a hierarchical LSTM to jointly associate shared semantics in words and regions. Recently, attention has become the most effective method to many-to-many approach as it enables to focus more on fragments relevant to shared semantics, and less on irrelevant fragments, where fragments can be either regions or words. Many attention modules have been proposed \cite{huang2017instance, lee2018stacked, huang2019bi}. sm-LSTM \cite{ huang2017instance} employs attention to sequentially find all possible shared semantics and simply associates them by optimizing pair-wise similarities. He et.al \cite{lee2018stacked} extend this idea and yield appealing performance. A stacked cross attention is designed to dynamically associate shared semantics by attending to words with respect to regions or attending to regions with respect to words, which makes many-to-many matching more interpretable as it changes the importance of target fragments based on fragments from another modality. Similar works are proposed in \cite{huang2019bi}. The attention mechanism they employ is within a fixed pattern, where the representation of shared semantics is a weighted combination over all the image regions or text words according to their contribution to shared semantics. However, only a fraction of regions or words relevant to shared semantics, integrating all of them will disturb the target semantic and thus lead to semantic misalignment. In this work, we address this issue by proposing a novel focal attention that eliminates irrelevant regions/words from shared semantics.

\section{Method}
The overall framework of our BFAN is illustrated in Figure 2. It consists of three components: feature extraction, focal attention and objective function. In this section, we first summarize the general attention framework in image-text matching, analyzing the semantic misalignment problem caused by existing framework in section 3.1. Then, we introduce our proposed focal attention and how to employ it into text-to-image and image-to-text matching, describing why and how to integrate them together in section 3.2. Last, we detail the objective function and feature extraction of our BFAN in section 3.3 and 3.4, respectively.

\subsection{General Attention Framework}
Without loss of generality, given an image-text pair consists of \begin{math}m\end{math} text words and \begin{math}n\end{math} image regions, a general image-text matching is to first project each image region and text word into a common \begin{math}d\end{math}-dimensional space using deep neural networks, getting text representation \begin{math} u\in \mathbb{R}^{m\times d} \end{math} and image representation \begin{math} v\in \mathbb{R}^{n\times d} \end{math}, and then learn to associate shared semantics in image and text using neural network blocks. The shared semantics are composed of multiple local shared semantics, such as local regions and words, and thus the overall objective is to maximumly improve the relevance of each local shared semantic:
\begin{equation}
R(u,v)=\frac{1}{K}\sum_{k}R(S_{k}^{u}, S_{k}^{v})
\end{equation}
where \begin{math}S_{k}^{u}\end{math} and \begin{math}S_{k}^{v}\end{math} denote \begin{math}k\end{math}-th shared semantic selected from image and text, respectively. \begin{math}R(\cdot)\end{math} denotes the relevance of shared semantics, which is computed using a similarity metric. \begin{math}K\end{math} denotes the number of shared semantics.

Existing attention methods find shared semantics by learning the attention distribution over all the fragments, e.g. regions or words, where fragments relevant to shared semantics obtain higher attention than others. Then, all the fragments are aggregated to represent shared semantics using a weighted combination:
\begin{equation}
S_{k}^{u}=\sum_{i=1}^{m}w_{ik}u_{i}, S_{k}^{v}=\sum_{j=1}^{n}w_{jk}v_{j}
\end{equation}
where \begin{math}w_{ik}\end{math} and \begin{math}w_{jk}\end{math} are attention distribution with respect to \begin{math}k\end{math}-th shared semantic, \begin{math}u_{i}\end{math} and \begin{math}v_{j}\end{math} denote \begin{math}i\end{math}-th word and \begin{math}j\end{math}-th region, respectively. However, not all the fragments support the specific shared semantic as many of them are irrelevant to it, the shared semantic will be more or less disturbed by irrelevant fragments if they are aggregated. More seriously, it will lead to semantic misalignment since different semantics cannot be appropriately decoupled. Therefore, it is necessary to represent the shared semantics by integrating a subset of fragments that are relevant to the target semantic.

\subsection{Our Focal Attention}
To address the semantic misalignment problem caused by the general attention framework, our focal attention proposes to learn a scoring function \begin{math} F\end{math} to identify fragments relevant to shared semantics, through which irrelevant fragments can be removed from shared semantics. Here, we set fragments with scores greater than zero as relevant, that is:
\begin{equation}
H(x)=\mathbb I(F(x)>0)
\end{equation}
where \begin{math}\mathbb I(\cdot)\end{math} is an indicator function, \begin{math}x\end{math} can be either regions or words.

It is impractical to find a fixed margin between relevant and irrelevant fragments based on the absolute value, e.g. similarity value between fragments and shared semantics, because it depends on iteratively updated fragment representations. Some attention approaches \cite{Luong2015Effective} attend to local fragments by simply masking fragments based on their position, but there is no connection with semantic relevance and fragment position. Inspired by non-local blocks proposed in \cite{wang2018non}, we determine the relevance of fragments by computing the relative importance of them to other fragments. The intuition behind this operation is irrelevant fragments always obtain low importance to the shared semantic compared with other relevant fragments. The scoring function is formulated as: 
\begin{equation}
F(x_{i})=\sum_{\forall x}f(x_{i},x_{j})g(x_{j})
\end{equation}

The pairwise function \begin{math}f(x_{i},x_{j})\end{math} computes relative importance of target \begin{math}i\end{math}-th fragment to \begin{math}j\end{math}-th fragment, and \begin{math}g(x_{j})\end{math} denotes the confidence of the fragment being compared, followed by an operation that sums up the weighted comparison results with all the other fragments. A fragment can be considered as relevant if it is similar to other relevant fragments with high confidence scores. Then, the \begin{math}k\end{math}-th shared semantic can be simply defined as:
\begin{equation}
S_{k}^{x}=\sum_{\forall x}w_{ik}x_{i}H(x_{i})
\end{equation}

In this work, our goal is to eliminate irrelevant fragments from context, which is totally different from traditional attention methods that focus on learning attention weight. In addition, differs from hard attention \cite{xu2015show} that estimates gradient using random sampling, the focal attention can compute gradient directly, because fragment except for irrelevant ones contribute to the forward-propagation. This allows for training the network both efficiently and effectively. We will depict how to employ the focal attention into text-to-image and image-to-text matching in 3.2.1 and 3.2.2.

\subsubsection{Text-to-Image Focal Attention}
In this work, we find shared semantics in image and text by fixing one modality and finding relevant fragments from another modality, where fragments in fixed modality are considered as the shared semantic. For text-to-image direction, text words are fixed as the shared semantic, we need to find relevant image regions for each text word. The overall framework includes three steps: preassign attention, identify relevant regions and reassign attention. To be specific, we first preassign attention score for each region, it is implemented by computing cosine similarities between regions and words, and normalizing them using softmax activation: 
\begin{equation}
w_{ij}=\sigma (\alpha \frac{u_{i}^{T}v_{j}}{\left \| u_{i} \right \|\left \| v_{j} \right \|}),i\in [1,m], j\in [1,n].
\end{equation}
where \begin{math}\sigma\end{math} denotes softmax activation. \begin{math}\alpha\end{math} is a scaling factor to further increase the gap between relevant and irrelevant regions, which is set as 20 in our implementation. 

Second, we identify relevant regions by scoring each region based on its allocated attention relative to other regions, regions with scores greater than zero are relevant regions, otherwise irrelevant.
\begin{equation}
F(v_{ij})=\sum_{t=1}^{n}f(v_{ij},v_{it})g(v_{it})
\end{equation}
We set \begin{math} f(v_{ij},v_{it})\end{math} as the difference of their preassigned attention to determine the relative attention of \begin{math}j\end{math}-th region to \begin{math}t\end{math}-th region since they are scalar value. The confidence score for \begin{math}t\end{math}-th region being compared is set as its relevance to the \begin{math}i\end{math}-th query word, e.g. \begin{math} \sqrt{w_{it}}\end{math}. Alternatively, we also set confidence of each image region as equal, it also proves to be effective in section 4.3.

After that, relevant regions can be selected by element-wise product between each image region and function \begin{math}H\end{math}. Third, we reassign attention weights for these selected relevant regions by renormalization. Note that irrelevant regions will not contribute to this process as their scores are zero:
\begin{equation}
w_{ij}^{'}=\frac{w_{ij}H(v_{ij})}{\sum_{j=1}^{n} w_{{ij}}H(v_{ij})}. 
\end{equation}

The reassigned attention weights will replace conventional attention \begin{math}w_{ik}\end{math} in equation 2, which allows to focus on most relevant regions as attention weights for irrelevant regions are zero. The shared semantic selected from the image based on \begin{math}i\end{math}-th word is computed as \begin{math}v_{i}^{'}=\sum_{j=1}^{n}w_{ij}^{'}v_{j}\end{math}. The global relevance of image and text is formulated as: 
\begin{equation}
R(u,v)=\frac{1}{m}\sum_{i=1}^{m}R(u_{i},v_{i}^{'}).
\end{equation}

\begin{table*}[]
	\setlength{\abovecaptionskip}{0pt}%
	\setlength{\belowcaptionskip}{2pt}%
	\captionsetup{width=.8\textwidth}
	\caption{Comparison results with baselines on Flickr30K. Image-to-Text denotes retrieve texts using image query, and Text-to-Image denotes retrieve images using text query. The best results are in bold. }
	\begin{tabular}{lccccccc}
		\hline
		\multicolumn{1}{c}{} & \multicolumn{3}{c}{Image-to-Text} & \multicolumn{3}{c}{Text-to-Image} &  \\
		Method & Recall@1 & Recall@5 & rmean & Recall@1 & Recall@5 & rmean & rsum \\ \hline
		Deep Fragment (single) \cite{DBLP:journals/corr/KarpathyJF14} & 16.4 & 40.2 & 28.3 & 10.3 & 31.4 & 20.9 & 98.3 \\
		HM-LSTM (single) \cite{Niu2017HierarchicalML} & 38.1 & - & 38.1 & 27.7 & - & 27.7 & 65.8 \\
		sm-LSTM (ensemble) \cite{huang2017instance} & 42.5 & 71.9 & 57.2 & 30.2 & 60.4 & 45.3 & 205.0 \\
		BSSAN (single) \cite{huang2019bi} & 44.6 & 74.9 & 59.8 & 33.2 & 62.6 & 47.9 & 215.3 \\
		VSE++ (single) \cite{Faghri2017VSE} & 52.9 & - & 52.9 & 39.6 & - & 39.6 & 92.5 \\
		DANs (single) \cite{nam2017dual} & 55.0 & 81.8 & 68.4 & 39.4 & 69.2 & 54.3 & 245.4 \\
		SCO (single) \cite{huang2018learning} & 55.5 & 82.0 & 68.8 & 41.1 & 70.5 & 55.8 & 249.1 \\
		SCAN (single) \cite{lee2018stacked} & 67.9 & 89.0 & 78.5 & 43.9 & 74.2 & 59.1 & 275.0 \\
		SCAN (ensemble) & 67.4 & 90.3 & 78.9 & 48.6 & 77.7 & 63.2 & 284.0 \\ \hline
		Ours: & \multicolumn{1}{l}{} & \multicolumn{1}{l}{} & \multicolumn{1}{l}{} & \multicolumn{1}{l}{} & \multicolumn{1}{l}{} & \multicolumn{1}{l}{} & \multicolumn{1}{l}{} \\
		BFAN-prob (single) & 65.5 & 89.4 & 77.5 & 47.9 & 77.6 & 62.8 & 280.4 \\
		BFAN-equal (single) & 64.5 & 89.7 & 77.1 & 48.8 & 77.3 & 63.1 & 280.3 \\
		BFAN-prob+equal (ensemble) & \textbf{68.1} & \textbf{91.4} & \textbf{79.8} & \textbf{50.8} & \textbf{78.4} & \textbf{64.6} & \textbf{288.7} \\ \hline
	\end{tabular}
\end{table*}
\begin{table*}[]
	\setlength{\abovecaptionskip}{0pt}%
	\setlength{\belowcaptionskip}{2pt}%
	\captionsetup{width=.8\textwidth}
	\caption{Comparison results with baselines on MSCOCO. Image-to-Text denotes retrieve texts using image query, and Text-to-Image denotes retrieve images using text query. The best results are in bold. }
	\begin{tabular}{lccccccc}
		\hline
		\multicolumn{1}{c}{} & \multicolumn{3}{c}{Image-to-Text} & \multicolumn{3}{c}{Text-to-Image} &  \\
		Method & Recall@1 & Recall@5 & rmean & Recall@1 & Recall@5 & rmean & rsum \\ \hline
		HM-LSTM (single) \cite{Niu2017HierarchicalML} & 43.9 & - & 43.9 & 36.1 & - & 36.1 & 80.0 \\
		sm-LSTM (ensemble) \cite{huang2017instance} & 53.2 & 83.1 & 68.2 & 40.7 & 75.8 & 58.3 & 252.8 \\
		BSSAN (single) \cite{huang2019bi} & 56.0 & 82.6 & 69.3 & 41.8 & 76.7 & 59.3 & 257.1 \\
		VSE++ (single) \cite{Faghri2017VSE} & 64.6 & - & 64.6 & 52.0 & - & 52.0 & 116.6 \\
		GXN (single) \cite{Gu2017Look} & 68.5 & - & 68.5 & 56.6 & - & 56.6 & 125.1 \\
		SCO (single) \cite{huang2018learning} & 69.9 & 92.9 & 81.4 & 56.7 & 87.5 & 72.1 & 307.0 \\
		SCAN (single) \cite{lee2018stacked} & 70.9 & 94.5 & 82.7 & 56.4 & 87.0 & 71.7 & \multicolumn{1}{l}{308.8} \\
		SCAN (ensemble) & 72.7 & 94.8 & 83.8 & 58.8 & 88.4 & 73.6 & 314.7 \\ \hline
		Ours: &  &  &  &  &  &  &  \\
		BFAN-prob (single) & 73.0 & 94.8 & 83.9 & 58.0 & 87.6 & 72.8 & 313.4 \\
		BFAN-equal (single) & 73.7 & 94.9 & 84.3 & 58.3 & 87.5 & 72.9 & 314.4 \\
		BFAN-prob+equal (ensemble) & \textbf{74.9} & \textbf{95.2} & \textbf{85.1} & \textbf{59.4} & \textbf{88.4} & \textbf{73.9} & \textbf{317.9} \\ \hline
	\end{tabular}
\end{table*}
  
\subsubsection{Image-to-Text Focal Attention}
Analogously, image regions are fixed as the shared semantic in image-to-text direction, we need to find relevant text words for each region. To this end, we first preassign attention by computing the similarity score between each image region and text word using cosine similarity, and normalize similarity scores of each word with respect to query region into [0,1] using softmax, attention on \begin{math}i\end{math}-th word is denoted as \begin{math} w_{ji}\end{math}. During this process, relevant words can be paid more attention, but irrelevant words also contribute to shared semantics between image region and target text. The second step is to score each word based on its preassigned attention relative to other words, that is: 
\begin{equation}
F(u_{ji})=\sum_{t=1}^{m}f(u_{ji},u_{jt})g(u_{jt})
\end{equation}

Next, the indicator function \begin{math}H(u_{ji})\end{math} is applied to identify relevant words based on computed score, where relevant words are set as one, otherwise zero. The attention for relevant words will be reassigned 
\begin{equation}
w_{ji}^{'}=\frac{w_{ji}H(u_{ji})}{\sum_{i=1}^{m} w_{ji}H(u_{ji})}. 
\end{equation}

The reassigned attention will be paid to all relevant words using element-wise product with their representations in \begin{math}d\end{math}-dimensional space. The shared semantic with \begin{math}j\end{math}-th region is selected from the text, computed as a weighted combination of relevant words \begin{math}u_{j}^{'}=\sum_{i=1}^{m}w_{ji}^{'}u_{i}\end{math}, where the learned focal attention determines the weight. The local relevance score \begin{math}R(v_{j},u_{j}^{'})\end{math} can be computed through cosine similarity. The global relevance score for image and text is calculated as the averaging of local relevance scores, that is:
\begin{equation}
R(v,u)=\frac{1}{n}\sum_{j=1}^{n}R(v_{j},u_{j}^{'}).
\end{equation}

\subsubsection{Bidirectional Focal Attention}
Focal attention on text-to-image and image-to-text are independent modules, where text-to-image focal attention learns to pick out a subset of image regions that semantically similar to each word, and image-to-text focal attention learns to pick out a subset of text words that semantically similar to each region. If we employ the focal attention to one direction, it will result in the preference to long text or complex image. It is because long text or complex image contains more information, and thus is more possible to get high response to query. Therefore, we present to jointly apply focal attention in two directions by combining their relevance as the overall relevance score. The bidirectional network will maximumly associate image-text pairs as it considers the semantic overlap instead of intersection in image and text. Specifically, we compute global relevance score between image and text in two directions separately, and then integrate them by taking their sum as the final score of image-text pairs, such that both directions contribute to the final relevance score:
\begin{equation}
S_{uv}=R(u,v)+R(v,u).
\end{equation}

\begin{table*}[]
	\setlength{\abovecaptionskip}{0pt}%
	\setlength{\belowcaptionskip}{2pt}%
	\caption{Ablation studies on Flickr30K, the best results are in bold.}
	\begin{tabular}{lccccccc}
		\hline
		\multicolumn{1}{c}{} & \multicolumn{3}{c}{Image-to-Text} & \multicolumn{3}{c}{Text-to-Image} &  \\
		Method & Recall@1 & Recall@5 & rmean & Recall@1 & Recall@5 & rmean & rsum \\ \hline
		BFAN-w/o-t2i & 60.4 & 85.4 & 72.9 & 46.3 & 76.5 & 61.4 & 268.6 \\
		BFAN-w/o-i2t & 63.0 & 87.2 & 75.1 & 45.9 & 75.0 & 60.5 & \multicolumn{1}{l}{271.1} \\
		BFAN-w/o-focal & 63.2 & 88.8 & 76.0 & 48.7 & 76.9 & 62.8 & 277.6 \\
		BFAN-prob & 65.5 & 89.4 & 77.5 & 47.9 & 77.6 & 62.8 & 280.4 \\
		BFAN-equal & 64.5 & 89.7 & 77.1 & 48.8 & 77.3 & 63.1 & 280.3 \\
		BFAN-prob+equal & \textbf{68.1} & \textbf{91.4} & \textbf{79.8} & \textbf{50.8} & \textbf{78.4} & \textbf{64.6} & \textbf{288.7} \\ \hline
	\end{tabular}
\end{table*}

\begin{table*}[]
	\setlength{\abovecaptionskip}{0pt}%
	\setlength{\belowcaptionskip}{2pt}%
	\caption{Ablation studies on MSCOCO, the best results are in bold.}
	\begin{tabular}{lccccccc}
		\hline
		\multicolumn{1}{c}{} & \multicolumn{3}{c}{Image-to-Text} & \multicolumn{3}{c}{Text-to-Image} &  \\
		Method & Recall@1 & Recall@5 & rmean & Recall@1 & Recall@5 & rmean & rsum \\ \hline
		BFAN-w/o-focal & 65.8 & 91.9 & 78.9 & 43.6 & 79.3 & 61.5 & 280.6 \\
		BFAN-w/o-i2t & 69.3 & 93.7 & 81.5 & 55.2 & 85.6 & 70.4 & \multicolumn{1}{l}{303.8} \\
		BFAN-w/o-t2i & 70.3 & 93.9 & 82.1 & 55.6 & 86.5 & 71.1 & 306.3 \\
		BFAN-prob & 73.0 & 94.8 & 83.9 & 58.0 & 87.6 & 72.8 & 313.4 \\
		BFAN-equal & 73.7 & 94.9 & 84.3 & 58.3 & 87.5 & 72.9 & 314.4 \\
		BFAN-prob+equal & \textbf{74.9} & \textbf{95.2} & \textbf{85.1} & \textbf{59.4} & \textbf{88.4} & \textbf{73.9} & \textbf{317.9} \\ \hline
	\end{tabular}
\end{table*}

Despite that recent approach \cite{huang2019bi} also employ bidirectional attention, ours are totally different. It derives from they simultaneously restraint scores at each direction while we restraint the overall score. This can relax constraint and avoid overfitting, because similar samples (one of them is long/complex) cannot be distinguished in single direction, it is inappropriate to restraint each direction.

\subsection{Objective Function}
To optimize the proposed network, we employ a structured ranking loss as the objective function, which has been proven to be able to maximize relevance scores of relevant image-text pairs and minimize that of irrelevant text-image pairs. Motivated by previous approach proposed by \cite{Faghri2017VSE}, we focus on hard negatives in each mini-batch, which produces maximum relevance score over any other irrelevant pairs. Given a pair of relevant image-text, we denote their relevance score as \begin{math}S_{uv}\end{math}, \begin{math}\bar{v} =\mathop{\arg\max}_{t\neq v} S_{ut}\end{math} denotes the hard negative when using the text to retrieve image, and \begin{math}\bar{u} =\mathop{\arg\max}_{t\neq u} S_{tv}\end{math} denotes the hard negative when using the image to retrieve text, their relevance score with the text or image are forced to be lower than that between relevant image-text pairs by a fixed margin, i.e.
\begin{equation}
L=[\alpha -S_{uv}+S_{\bar{u}v}]_{+}+[\alpha -S_{uv}+S_{u\bar{v}}]_{+}.
\end{equation}
where, \begin{math}[x]_{+}\equiv max(x,0)\end{math}, we set the loss as zero if relevance score with hard negative is not as large as that with relevant pairs. The margin \begin{math}\alpha\end{math} is a hyperparameter that is set as 0.2.

\subsection{Feature Extraction}
\subsubsection{Image Feature}
In many-to-many image-text matching, each image is comprised of multiple regions. We detect salient regions that contribute most to the global semantic, and encode each of them into feature vectors. In this work, we detect salient regions using a popular object detection tool Faster R-CNN \cite{Ren2017Faster}. The tool predicts object bounding boxes and scores them. We select top K (K=36) salient objects according to their scores, and extract mean-pooled convolutional features for these bounding boxes using pretrained ResNet-101 \cite{He2016DeepRL}. A fully-connected layer is applied to transform features into target \begin{math}d\end{math}-dimensional feature vector.
\subsubsection{Text Feature}
Similar to the image, each text contains a set of words, we encode each word into \begin{math}d\end{math}-dimensional feature vectors as well as image region and combine them as the global feature of the text. To this end, we employ the bidirectional GRU to integrate the feedforward and backward contextual information into word representations. Specifically, we first split a text into multiple words, and embed each word into a low-dimensional vector to decrease the computation cost of GRU, which are then fed into bidirectional GRU. After multi-step iterations, the average of forward and backward hidden state can be considered as the text representation, which contains \begin{math}d\end{math}-dimensional features for each word in the text.

\section{Experiments}
\subsection{Experimental Setup}
\subsubsection{Datasets}We conduct several experiments on image-text matching benchmarks, Flickr30K \cite{Plummer2017Flickr30k} and MSCOCO \cite{Lin2014Microsoft}. Flickr30K is a standard dataset for image-text matching, it contains 31,000 images and 155,000 texts in total, each image relates to five texts. Following \cite{Mao2015Deep,DBLP:journals/corr/KarpathyJF14,Liu2017LearningAR}, we split Flickr30K benchmark into 29K training images, 1K validation images and 1K testing images. MSCOCO is a large-scale benchmark that contains 123,287 images with five texts each. We use 113,287 images for training, 5,000 images for validation and 5,000 for testing follow \cite{lee2018stacked,Faghri2017VSE}. We report experimental results through averaging 5-folds on 1K test images.

\begin{figure*}[!t]
	\centering
	\includegraphics[width=0.95\linewidth]{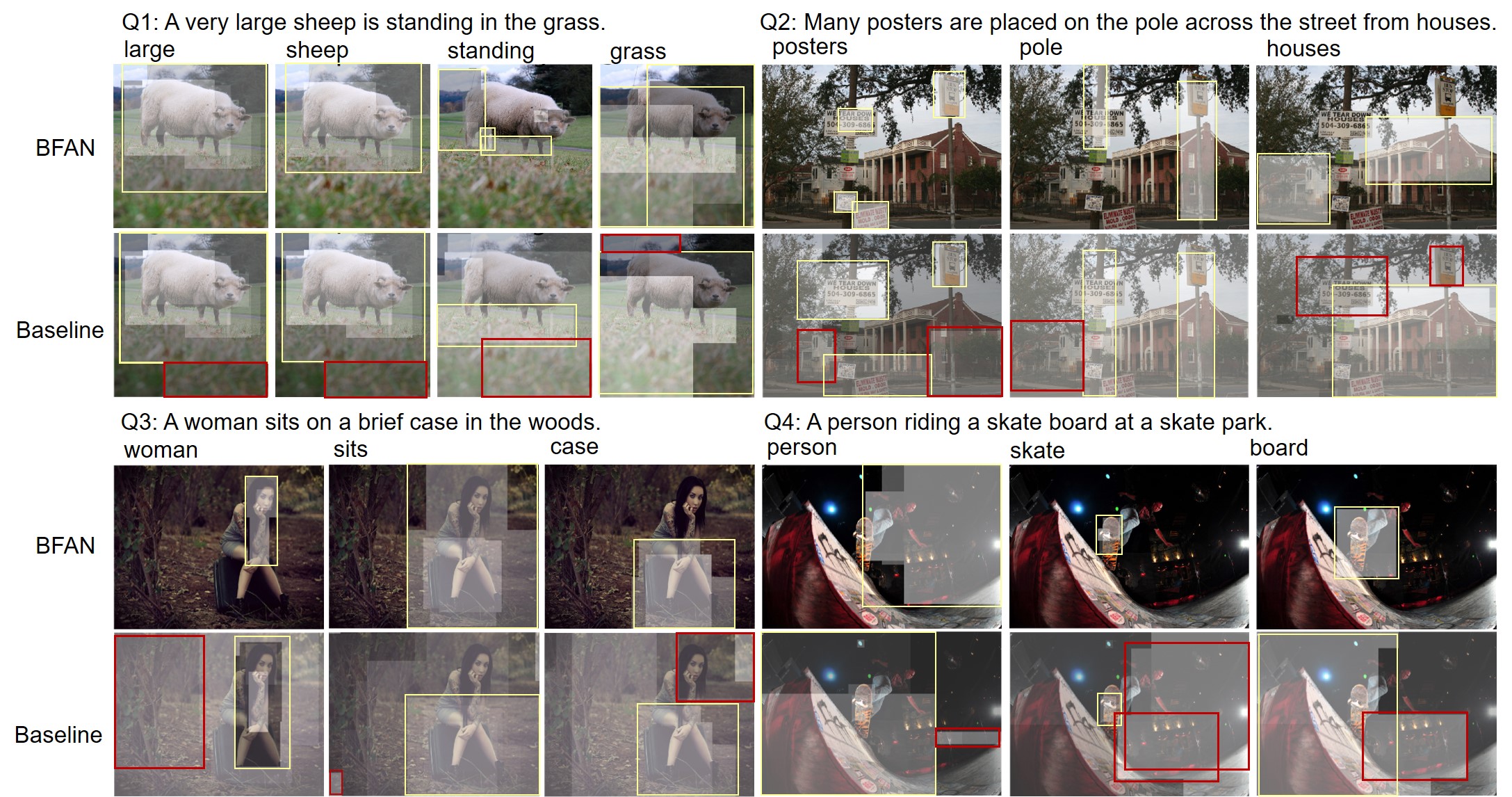}
	\caption{Visualization of our focal attention and conventional attention \cite{lee2018stacked} with respect to each word shown at the top left corner of each image, where brighter regions obtain more attention. Relevant and irrelevant regions are outlined in yellow and red boxes, respectively, which shows our attention always focus on relevant regions while \cite{lee2018stacked} distracts attention as it attends to many irrelevant regions in addition to relevant ones.}
\end{figure*}

\subsubsection{Evaluation}We evaluate the performance of our proposed approach by reporting Recall@K (K = 1,5) values for both image-to-text and text-to-image matching task. The Recall computes the proportion of correct image or text being retrieved among top K results. In addition, we compute mean value of Recall (rmean) in each direction, and sum of Recall (rsum) to show overall performance.

\subsubsection{Settings} The proposed network is implemented using PyTorch, and trained on 1 NVIDIA TITAN Xp optimized by Adam. We start training the network with learning rate 0.0002 on Flickr30K and 0.0005 on MSCOCO, and decay by 0.1 after every 10 epochs. The mini-batch size is set as 32. Our network requires to train 15 epochs on Flickr30K and 20 epochs on MSCOCO, training instances are randomly shuffled at each epoch. We set the dimensionality of image region representations as 1024. The initial one-hot vector of word embeddings are covert to 300-dimensional, and then fed into bidirectional GRU that produces 1024-dimensional representations. 


\subsubsection{Baselines}We select most representative works as baselines, including the first many-to-many approach Deep Fragment \cite{DBLP:journals/corr/KarpathyJF14}, recent works Hierarchical Multimodal LSTM (HM-LSTM) \cite{Niu2017HierarchicalML}, Selective Multimodal LSTM (sm-LSTM) \cite{huang2017instance}, Bi-Directional Spatial-Semantic Attention Networks (BSSAN) \cite{huang2019bi} and state-of-the-art Stacked Cross Attention Network (SCAN) \cite{lee2018stacked}. We also make comparisons with most recent one-to-one matching works, including Dual Attention Networks (DANs) \cite{nam2017dual}, visual-semantic embeddings (VSE++) \cite{Faghri2017VSE}, semantic-enhanced image and sentence matching model (SCO) \cite{huang2018learning} and generative cross-modal feature learning framework (GXN) \cite{Gu2017Look}. We provide two versions of focal attention implemention, including BFAN-prob and BFAN-equal, where one takes confidence of each compared fragment into account, and another one treats each fragment equally. Note that some approaches use ensemble model by averaging the global relevance score of two single models, we also provide single and ensemble model to make a fair comparison, it is achieved by averaging relevance scores calculated by single models. 

\subsection{Comparison Results}
We conduct extensive experiments on Flickr30K and MSCOCO, respectively. Quantitative results on Flickr30K are listed in Table 1. In real application, top-1 result is more concerned by users, so improving Recall@1 is crucial to improve user experience, this is exactly advantage of focal attention. It is observed that our approach achieves more improvement on Recall@1 than other metrics. The BFAN achieve 68.1\% and 50.8\% Recall@1 value on image-to-text and text-to-image matching, respectively. It is the first time that Recall@1 in text-to-image matching over 50\% on Flickr30K benchmark, getting a 2.2\% relative improvement than state-of-the-art SCAN. Compared with VSE++, which also optimizes on hard negatives, we can obtain relative Recall@1 gains with 15.2\% and 11.2\%. Although BSSAN proposes similar bidirectional networks, our BFAN outperforms it with over 18\% relative gains on average since relevant fragments are tightly correlated without interference of irrelevant ones. Compared with two most effective one-to-one methods, SCO and GXN, our approach not only outperforms them, but also learns more fine-grained region-word correspondence, which is significant for real multimodal application.

Quantitative results on MSCOCO are listed in Table 2. MSCOCO is a larger image-text matching benchmark, our improvement on MSCOCO shows the proposed approach has excellent and stable capability of generalization. Our single model outperforms state-of-the-art single model with a relative 5.3\% \begin{math}\sim\end{math}5.4\% gain in terms of rsum, our ensemble model also outperforms the best ensemble model. Note that our BFAN achieves more improvement on Recall@1, which is significant for image-text matching.

\subsection{Ablation study}
Table 3 shows ablation study results on Flickr30K. Both focal attention and bidirectional version contribute towards the overall performance. To evaluate the effect of focal attention, we remove focal attention in our full model, and employ traditional attention on both image-to-text and text-to-image directions, denoted as BFAN-w/o-focal. Focal attention proves to be critical to improving overall matching performance, especially for Recall@1 as it removes most irrelevant fragments. To evaluate the effect of bidirectional focal attention, we employ focal attention in either image-to-text or text-to-image direction, referred as BFAN-w/o-t2i and BFAN-w/o-i2t. Results show that the single directional focal attention will decrease all the Recall value by nearly 2\% on average compared with full single model. It derives from the single model is partial to long text and complex image as they are more likely to contain target fragments. Our bidirectional attention avoids this by considering the proportion of relevant fragments instead of their absolute quantity. In addition, we also investigate the effect of our focal attention with different implementation, i.e. BFAN-prob and BFAN-equal. Both of them can achieve great performance, and combining them can largely improve the Recall value. It is because their combination can learn a better mode of relevant fragments selection. Ablation study results on MSCOCO benchmark is shown in Table 4. Different from the results on Flickr30K, BFAN-w/o-focal achieves better performance than BFAN-w/o-t2i and BFAN-w/o-i2t, the two full single models still outperform other models at all the evaluation metrics, which shows focal attention and bidirectional version complement to each other, and can be stably applied to different datasets.

\subsection{Attention Visualization} To better understanding the difference in focal attention and conventional attention, we visualize attention weights for each image region with respect to query word in Figure.3. We make comparisons with the baseline model \cite{lee2018stacked}, attention weight of each bounding box (image region) released by bottom-up attention \cite{anderson2018bottom} is computed using BFAN and baseline, respectively. We use the brightness to visualize attention weights, and brighter regions obtain more attention. We show attention distribution in image regions with respect to nouns and verbs in the text, where the chosen word is shown at the top left corner of each image. Relevant and irrelevant regions are mainly outlined in yellow and red boxes. It is observed that BFAN learns better semantic alignment. For example, in Q1, ``Standing'' corresponds to the gigot by our BFAN, while baseline also aligns it with irrelevant regions, like sky and grass.

\subsection{Qualitative Results} We also provide visualization for text-to-image and image-to-text matching. For text-to-image matching shown in Figure 4, we show top 3 ranked images for each text query. Images in first three columns are retrieved by our approach, and the last three columns are by baseline \cite{lee2018stacked}. The correctly retrieved images are in green box. Long and short text queries can be well matched with their most relevant images. For the first text query, the correct image ranks first by our BFAN despite local regions in the second image hit some keywords, such as ``black snow pants'' and ``wearing a black coat''. Baseline model gives incorrect ranks since irrelevant regions disturb the semantic alignment. For example, they will attend to irrelevant ``red coat'' when matching ``black coat'', which will lower the response to query word, but the incorrect image will give a high response since most people wear ``black coat''. 

\begin{figure}[htb]
	\centering
	\setlength{\abovecaptionskip}{1pt}%
	\setlength{\belowcaptionskip}{0pt}%
	\includegraphics[width=1\linewidth]{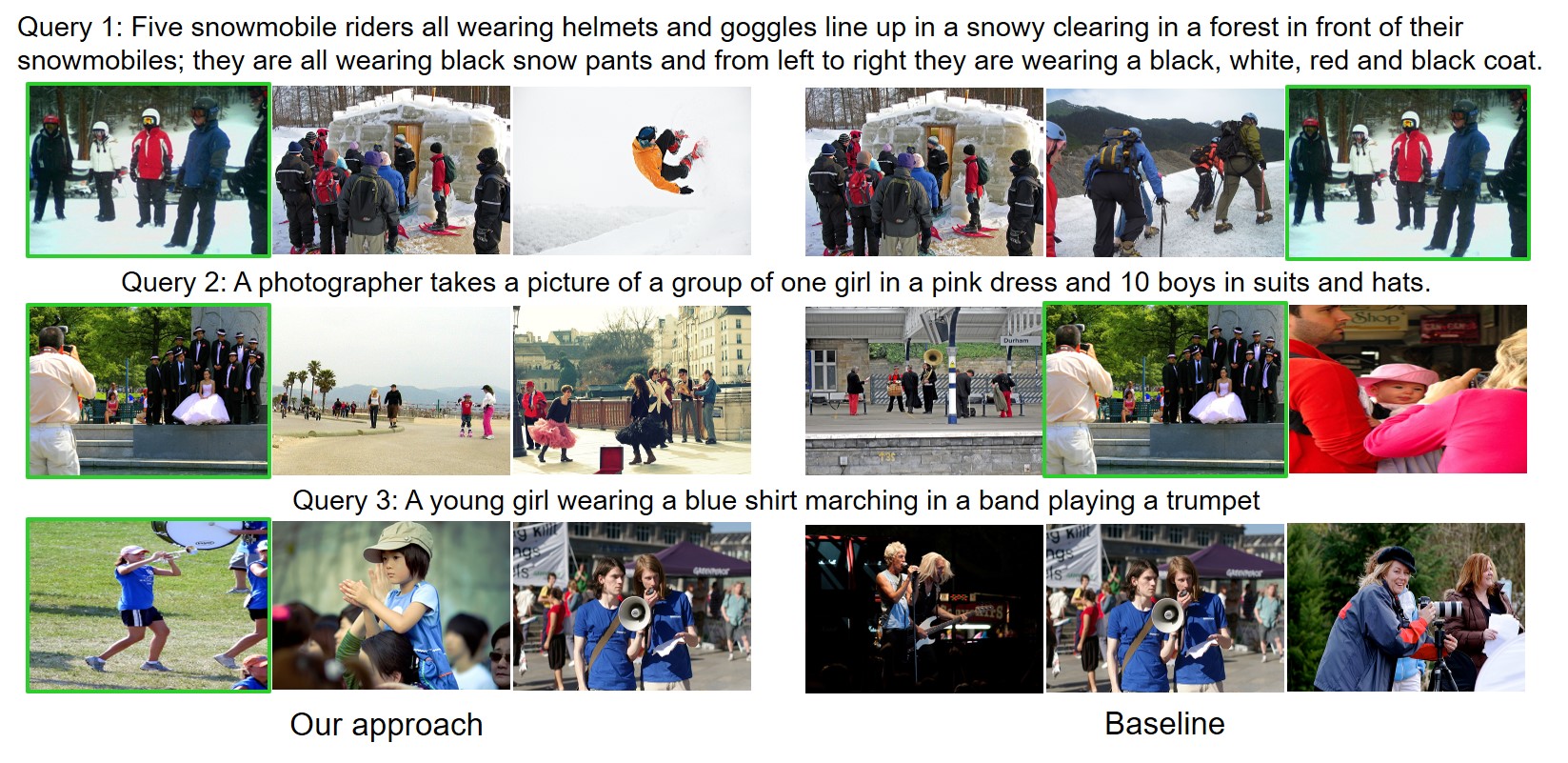}
	\caption{Text-to-image matching by our approach and baseline \cite{lee2018stacked}. For each text query, we list top-3 ranked images from left to right, where correct answers are outlined as green box. The first three columns are our results and the last three columns are baseline results.}
	\Description{t2i}
\end{figure}

\begin{figure}[htb]
	\centering
	\setlength{\abovecaptionskip}{2pt}%
	\setlength{\belowcaptionskip}{0pt}%
	\includegraphics[width=1\linewidth]{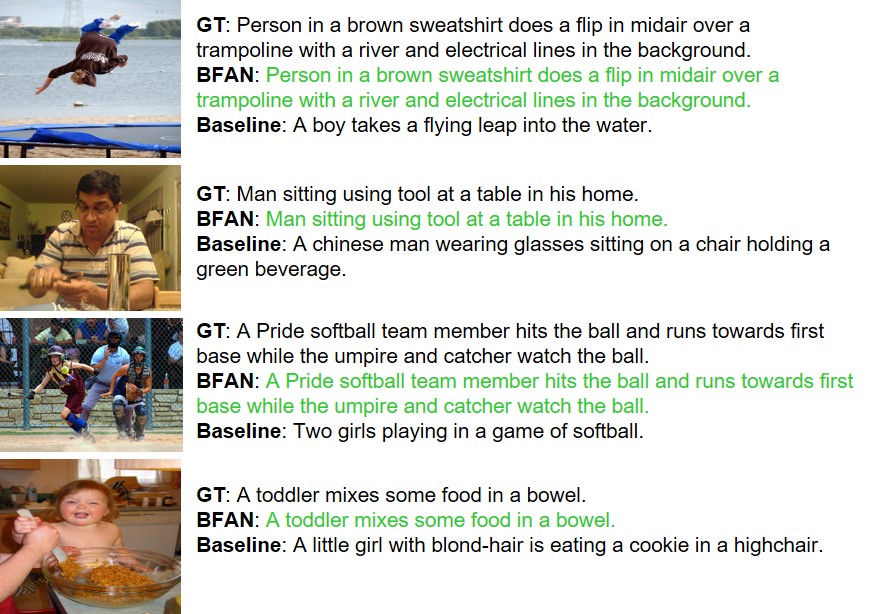}
	\caption{Image-to-text matching by our approach and baseline \cite{lee2018stacked}. For each image query, we provide the ground truth (GT), top-1 ranked text by BFAN and baseline at the right-hand of the image, where correct ones are marked as green.}
	\Description{i2t}
\end{figure}

We visualize image-text matching performance in Figure 5. The ground truth (GT), top-1 ranked text produced by our approach and baseline \cite{lee2018stacked} are listed at the right-hand of each image query, where correct results are marked as green. As shown in the first example, baseline gets an incorrect result as it always attends to keyword ``water'', and thus it plays an important role while querying other objects like the person and action ``flip''. It further confirms the necessity of using focal attention. Results also show that our approach can capture and discriminate more detailed information. For example, for the last example, baseline gets the wrong answer since it cannot identify ``cookie'' and ``highchair'' despite most other keywords match the query image, while the BFAN performs well.

\section{Conclusion}
In this paper, we propose a novel bidirectional focal attention model for image-text matching. Different from conventional attention, our focal attention only attends to fragments relevant to query fragment, which can address semantic misalignment caused by existing attention methods. The directional version can also avoid the preference to long text or complex image. We conduct comprehensive experiments that demonstrate the proposed method can significantly outperform state-of-the-art. Future research directions include applying the focal attention into other cross-modal applications such as translation, image caption and visual question answering.

\begin{acks}
	This work is supported by the National Natural Science Foundation
	of China, Grant No.61502477, the National Key R\&D Program with No.2016QY03D0503, 2016YFB081304, Strategic Priority Research Program of Chinese Academy of Sciences, Grant No.XDC02040400.
\end{acks}
\balance

%

%
\bibliographystyle{ACM-Reference-Format}
\bibliography{sample-base}

%
\appendix
\end{document}